\documentclass{article}
\usepackage[latin9]{inputenc}
\usepackage{geometry}
\usepackage{float}
\usepackage{booktabs}
\usepackage{amsmath}
\usepackage{amssymb}
\usepackage[unicode=true,pdfusetitle,
 bookmarks=true,bookmarksnumbered=false,bookmarksopen=false,
 breaklinks=false,pdfborder={0 0 1},colorlinks=false]
 {hyperref}
\usepackage{breakurl}

\makeatletter

\providecommand{\tabularnewline}{\\}

\usepackage{graphicx}
\usepackage{grffile}
\usepackage{siunitx}
\usepackage{multirow}
\usepackage{footnote}
\usepackage{breakurl}

\providecommand{\tabularnewline}{\\}

\usepackage{fullpage}\usepackage[justification=centering]{caption}\usepackage[all]{hypcap}\usepackage{microtype}

\makeatother

\begin{document}

\title{Ratio B(E2, 4$\rightarrow2$)/B(E2, 2$\rightarrow$0) in Even-Even
Nuclei: Apparent Anomalous Behavior of the Chromium Isotopes}

\author{Daniel Hertz-Kintish$^{1}$, Larry Zamick$^{1}$, Shadow JQ Robinson$^{2}$\\
 \textit{$^{1}$Department of Physics and Astronomy, Rutgers University,
Piscataway, New Jersey 08854, USA}\\
 \textit{$^{2}$Department of Physics, Millsaps College Jackson MS
39210} }

\date{\today}
\maketitle
\begin{abstract}
We consider the ratio RE4 = B(E2,$4\rightarrow2$)/B(E2,$2\rightarrow$0)
for the lowest $2^{+}$ and $4^{+}$ states in even-even nuclei. In
the rotational and vibrational models and the shell model calculations
here considered, RE4 is greater than one, however empirically, using
NNDC adopted half-lives and energies for $^{48}$Cr and $^{50}$Cr,
this ratio is less than one. 
\end{abstract}

\section{Introduction and Motivation}

In shell model calculations of B(E2)s it is necessary to use effective
charges if one hopes to get reasonable agreement with experiment.
A popular choice, but my no means the only one is 1.5 for the proton
and 0.5 for the neutron (as compared with the bare effective charges
of 1 and 0 respectively). One may also attempt to calculate these
effective charges from fundamentals. If for a given nucleus one gets
a good fit to B(E2, $2\rightarrow0$), one cannot regard this as an
impressive result. One has chosen the effective charge so that the
theory fits experiment---one parameter has been adjusted to fit one
experimental datum. \\

In this work we define REJ as B(E2, J $\rightarrow$ J-2)/ B(E2, 2
$\rightarrow$ 0). We will focus mainly on RE4$=$B(E2, $4\rightarrow2$)/B(E2,
$2\rightarrow0$), as there exists significantly more experimental
data on B(E2)s for $2^{+}$ and $4^{+}$ states than there is for
higher states. By so doing, to a large extent we take the effective
charge out of play. There is considerable empirical data in the National
Nuclear Data Center, NNDC, on half-lives of the lowest $2^{+}$ and
$4^{+}$ states of many even-even nuclei so that we can obtain RE4
empirically. \\

Before proceeding to the shell model we note that there are simple
formulas for this ratio in the rotational model and the vibrational
model.

\begin{equation}
RE4_{rotational}=10/7\approx1.429
\end{equation}
\begin{equation}
RE4_{vibrational}=2
\end{equation}

This comes from more detailed formulae of Bohr and Mottelson \cite{Bohr}:

\begin{equation}
REJ_{rotational}=\frac{\langle J\text{ }0\text{ }2\text{ }0|(J-2)\text{ }0\rangle}{\langle2\text{ }0\text{ }2\text{ }0|0\text{ }0\rangle}
\end{equation}
\begin{equation}
REJ_{vibrational}=J/2
\end{equation}

\section{Shell Model}

The main thrust of this work will be a comparison of the empirical
adopted values of RE4 from NNDC \cite{NNDC} against calculated values.
Besides the simple rotational and vibrational results provided in
the previous section we also consider the shell model calculations
of Robinson et al. \cite{R2005,R2014}. In the $f-p$ shell, the program {\sc NuShell} \cite{Nu} is used in the FPD6 and GXFP1 interactions for the lighter nuclei. The program {\sc ANTOINE} \cite{Ant} is used for the JJ45 and JUN45 interactions for the heavier nuclei. In Table 1, the empirical
results in the second column should be compared with the shell model
results using an FPD6 interaction in the third column. We will discuss
the fourth column later.

\begin{savenotes} 
\begin{table}[H]
\centering \protect\caption{The B(E2) ratio RE4 empirical vs. shell model calculations}

\begin{tabular}{rrrrr}
\toprule 
\addlinespace
\protect Nuclide  & {Empirical}  & FPD6  & T0FPD6\protect%
\footnote{T0FPD6 is derived from FPD6 by setting all the T=0 2-body matrix elements
of FPD6 to zero but keeping the T=1 elements unchanged.%
} & GXPF1 \tabularnewline
\addlinespace
$^{44}$Ti  & 2.269  & 1.362  & 1.083 & 1.291 \tabularnewline
$^{46}$Ti  & 1.033  & 1.419  & 1.153 & 1.264 \tabularnewline
$^{48}$Ti  & 1.254  & 1.519  & 1.192 & 1.414 \tabularnewline
$^{48}$Cr  & 0.834  & 1.396  & 1.289 & 1.351 \tabularnewline
$^{50}$Cr  & 0.7558  & 1.450 & 1.278 & 1.422 \tabularnewline
$^{52}$Fe  & 1.839  & 1.460 & 1.182 & 1.304\tabularnewline
\end{tabular}\label{table:nonlin} 
\end{table}

\end{savenotes} \raggedright

\begin{table}[H]
\centering \protect\caption{Quadrupole moments of 2$^{+}$/4$^{+}$}

\begin{tabular}{rrrrr}
\toprule 
\addlinespace
\protect Nuclide  & {Empirical}  & FPD6  & T0FPD6  & GXPF1 \tabularnewline
\addlinespace
$^{44}$Ti  &  & -21.572/-28.918  & -0.945/-7.391 & -5.132/-16.378 \tabularnewline
$^{46}$Ti  &  & -23.505/-31.02  & -8.777/-17.263 & -12.72/-22.985 \tabularnewline
$^{48}$Ti  &  & -18.807/-20.724  & -9.402/-8.767 & -13.719/-11.424 \tabularnewline
$^{48}$Cr  &  & -35.416/-45.472  & -22.825/-33.718 & -30.16/-40.424 \tabularnewline
$^{50}$Cr  &  & -32.914/-41.867 & -22.317/-31.931 & -28.34/-35.738 \tabularnewline
$^{52}$Fe  &  & -33.642/-38.486 & -25.686/-23.508 & -30.4509/-37.5636\tabularnewline
\end{tabular}\label{table:nonlin} 
\end{table}

\raggedright

The first thing to notice is that the shell model results all have
RE4 bigger than one; this is also the case for the rotational and
vibrational models. The shell model results are closer to the rotational
value of 10/7 than the large value of 2 from the vibrational model.
\\

The empirical results seem to fluctuate quite a lot from nucleus
to nucleus as compared with the steadier FPD6 shell model results.
Of particular note is the fact that for $^{48}$Cr and $^{50}$Cr,
the RE4 values are less than one---0.862 and 0.756 respectively. We
here do not attempt to explain this behavior but present it as a nuclear
structure puzzle. Using data from Brandolini, et al. \cite{Bran1,Bran2},
instead of data from NNDC, the values of the RE4 of $^{48}$Cr and
$^{50}$ Cr are 0.997 and 0.921 respectively. Although they are somewhat
larger, these values remain less than one and well below the shell
model results. \\

Also worth mentioning is the fact that for $^{44}$Ti we get a very
large RE4, even exceeding the vibrational value of 2.

In Table 2 we present the raw NNDC data---the excitation energies
of the lowest 2$^{+}$ and 4$^{+}$ states (E2 and E4) as well as
the ratio E4/E2. We also show the half lives of these states T$_{1/2}$.
We obtain RE4 from this data by the formula 
\begin{equation}
RE4=\frac{(T_{1/2})_{2^{+}}(E_{2^{+}})^{5}}{(T_{1/2})_{4^{+}}(E_{4^{+}}-E_{2^{+}})^{5}}
\end{equation}

\begin{table}[H]
\centering \protect\caption{NNDC adopted values of half-lives and energy levels used to obtain
RE4s}

\begin{tabular}{rrrrrr}
\toprule 
\addlinespace
Nuclide  & Yrast State  & Half-life (ps)  & Energy (keV)  & Empirical RE4  & Energy Ratio E4/E2\tabularnewline
\addlinespace
\multirow{2}{*}{$^{44}$Ti}  & 2$^{+}$  & 3.1  & 1083.06  & \multirow{2}{*}{2.285}  & \multirow{2}{*}{2.2643}\tabularnewline
 & 4$^{+}$  & 0.42  & 2452.33 &  & \tabularnewline
 &  &  &  &  & \tabularnewline
\multirow{2}{*}{$^{46}$Ti}  & 2$^{+}$  & 5.32 & 889.28 & \multirow{2}{*}{1.034}  & \multirow{2}{*}{2.2601}\tabularnewline
 & 4$^{+}$  & 1.62  & 2009.846 &  & \tabularnewline
 &  &  &  &  & \tabularnewline
\multirow{2}{*}{$^{48}$Ti}  & 2$^{+}$  & 4.04  & 983.539  & \multirow{2}{*}{1.255}  & \multirow{2}{*}{2.3341}\tabularnewline
 & 4$^{+}$  & 0.762  & 2295.654 &  & \tabularnewline
 &  &  &  &  & \tabularnewline
\multirow{2}{*}{$^{48}$Cr}  & 2$^{+}$  & 7.3  & 752.19  & \multirow{2}{*}{0.862}  & \multirow{2}{*}{2.4707}\tabularnewline
 & 4$^{+}$  & 1.23  & 1858.47 &  & \tabularnewline
 &  &  &  &  & \tabularnewline
\multirow{2}{*}{$^{50}$Cr}  & 2$^{+}$  & 9.08  & 783.32  & \multirow{2}{*}{0.756}  & \multirow{2}{*}{2.4017}\tabularnewline
 & 4$^{+}$  & 2.22  & 1881.31 &  & \tabularnewline
 &  &  &  &  & \tabularnewline
\multirow{2}{*}{$^{52}$Fe}  & 2$^{+}$  & 7.8  & 849.45  & \multirow{2}{*}{1.839}  & \multirow{2}{*}{2.8072}\tabularnewline
 & 4$^{+}$  & 0.22  & 2384.55 &  & \tabularnewline
\end{tabular}\label{table:nonlin} 
\end{table}

\raggedright

There are calculations in heavier nuclei by Robinson et al.\cite{R2014}
from which one can extract RE4 but there is as of yet no empirical
data for these nuclei. These results are presented in Table 4 for
the interactions JUN45 and JJ4B. The nuclei involved are $^{88}$Ru,$^{92}$Pd
and $^{96}$Cd. The calculated shapes are quite different, oblate,
near zero deformation, and prolate, respectively, however the calculated
values of RE4 are similar, all bigger than one and close to the rotational
limit.

\begin{savenotes} 
\begin{table}[H]
\centering \protect\caption{Calculated values of RE4 for heavier nuclei}

\begin{tabular}{rrrrr}
\toprule 
\addlinespace
Nuclide  & JUN45  & T0JUN45 & JJ4B  & T0JJ4B\tabularnewline
\addlinespace
$^{88}$Ru  & 1.554  & 1.314  & 1.458 & 1.416\tabularnewline
$^{92}$Pd  & 1.256  & 1.040  & 1.359 & 2.662\protect%
\footnote{Both values are extremely small, a factor of 200-300 x smaller than
the full values%
}\tabularnewline
$^{96}$Cd  & 1.356  & 1.102  & 1.330  & 1.296\tabularnewline
\end{tabular}\label{table:nonlin} 
\end{table}

\end{savenotes} \raggedright

\begin{savenotes} 
\begin{table}[H]
\centering \protect\caption{Quadrupole moments 2$^{+}$/4$^{+}$ for heavier nuclei}

\begin{tabular}{rrrrr}
\toprule 
\addlinespace
Nuclide  & JUN45  & T0JUN45 & JJ4B  & T0JJ4B\tabularnewline
\addlinespace
$^{88}$Ru  & 36.721/43.26  & 7.714/13.59  & 28.9297/36.513 & -5.85/-2.9049\tabularnewline
$^{92}$Pd  & -3.559/-7.968  & -3.778/4.929  & 4.5599/11.161 & 8.4856/-5.48\tabularnewline
$^{96}$Cd  & -19.296/-21.486  & -7.328/-12.778  & -16.4119/-15.208  & -15.175/-14.044\tabularnewline
\end{tabular}\label{table:nonlin} 
\end{table}

\end{savenotes} \raggedright

We last discuss the column T0FPD6 of Table 1. In their 2005 work Robinson
et al. \cite{R2005} wanted to see the effects of the T=0 interaction
on energy levels and B(E2)s. They did so by removing and then replacing
all the 2-body T=0 matrix elements. InT0FPD6 they kept the T=1 matrix
elements of FPD6 unchanged but set all the T=0 ones to zero. Focusing
here on RE4 we see that with T0FPD6 the values of RE4 are much closer
to one than with the full FPD6 interaction. We can thus say tha the
effect of the T=0 interaction is to make RE4 larger and close to the
rotational limit. 

We here add new results on static quadrupole moments of the lowest
2$^{+}$ and 4$^{+}$ states. These are shown in Tables 2 and 4. There
are no experiments to compare with. For the most part when the T=0
interaction is turned on the magnitudes of the quadrupole moments increase.
This is true both for prolate and oblate cases.

A note should be made on the rotational model. As J increases, REJ
also increases in this model. However, in the shell model, REJ ultimately
decreases. This is shown not only theoretically\cite{R2014} but also
in the works of Brandolini et al. where they determined several values
of B(E2, J $\rightarrow$ J-2) in $^{48}$Cr,$^{50}$Cr \cite{Bran1}and
$^{46}$Ti\cite{Bran2}. \\

Given the apparent anomalies for the Chromium nuclei, one can either
question the theory or one can question the experiment. At this point
we are unable to make a definitive statement in either direction, but
we feel it makes the most sense to first encourage more experiments
on these nuclei. It is possible that the technology has improved and
the relevant values can be measured more precisely. \\

Daniel Hertz-Kintish thanks the Rutgers Aresty Research Center for
Undergraduates for support during the 2014 summer session. 

\end{document}